# A method based on Generative Adversarial Networks for disentangling physical and chemical properties of stars in astronomical spectra

Raúl Santoveña[a], Carlos Dafonte[a], Minia Manteiga[b]

[a]*CIGUS CITIC - Department of Computer Science and Information Technologies, University of A Coruña, Campus de Elviña s/n, A Coruña, 15071, Spain*
[b]*CIGUS CITIC - Department of Nautical Science and Marine Engineering, University of A Coruña, Paseo de Ronda 51, A Coruña, 15011, Spain*

**Abstract**

Data compression techniques that are focused on information preservation have become essential in the modern era of big data. In this work, an encoder-decoder architecture has been designed where adversarial training, a modification of the traditional autoencoder, is used in the context of astrophysical spectral analysis. The goal of this proposal is to obtain an intermediate representation of the astronomical stellar spectra, in which the contribution to the flux of a star due to the most influential physical properties (its surface temperature and gravity) disappears and the variance reflects only the effect of the chemical composition over the spectrum. We apply a scheme of deep learning with the aim of unraveling in the latent space the desired parameters of the rest of the information contained in the data. This work proposes a version of adversarial training that makes use of one discriminator per parameter to be disentangled, thus avoiding the exponential combination that occurs in the use of a single discriminator, as a result of the discretization of the values to be untangled. To test the effectiveness of the method, synthetic astronomical data are used from the APOGEE and Gaia surveys. With our approach, a significant enhancement in disentangling is proven, reaching an improvement in the $R^2$ score values of up to 0.7. In conjunction with the work presented, an ad-hoc framework (GANDALF) is provided, which allows the replication, visualization, and extension of the method to domains of any nature.

# 1. Introduction

Finding representations of the data that can ease the extraction of useful information and improve algorithm performance in classification or parametrization problems has become a field in itself in the machine learning community, and is known as representation learning. Ideally, this task is deeply related to the goal of finding the factors of variation that can explain the original data. The process of unraveling these underlying factors in a comprehensive representation is called disentangled representation. There is abundant literature on the problem of how to decode or separate representations of a signal into projections that include only information relevant to a specific problem. It is a ubiquitous problem in any application aimed at the automated extraction of information from massive and complex data. In Wang et al. [1] the current state of the literature is exhaustively reviewed discussing different methodologies, metrics, models, and applications. A paradigmatic example is InfoGAN (Chen et al. [2]), a generative adversarial network that maximizes the mutual information between a small subset of latents (variables inferred by the model) and observations, which was successfully applied to several image recognition problems.

Since they were first proposed, Generative Adversarial Networks (GANs) (Goodfellow et al. [3]) have become one of the most often used methods to learn disentangled representations in a variety of problems. Conditional GANs (Mirza and Osindero [4]) were quickly introduced, modifying the original architecture by feeding the data with some additional labels, which allows for the generation of data conditioned by attributes. In recent years, multiple alternatives have been proposed that study novel architectures and supervised and unsupervised methods, and apply them to all types of domains. One such alternative is the Fader Networks architecture (Lample et al. [5]), which proposes generative autoencoders as a method to isolate factors of variation in intermediate representations.

In the astrophysical domain, the problem of parameter disentangling, in this case applied to stellar spectra, was first tested by Price-Jones and Bovy [6], who fitted a polynomial model of the non-chemical parameters to every single wavelength bin in a grid of synthetic spectra. They used a clustering algorithm on a compressed representation of the residuals obtained after principal component analysis to identify chemically similar groups. Disen-tanglement using neural networks has been recently addressed in the works by de Mijolla et al. [7] [8]. In the first of them, a neural network with a supervised disentanglement loss term was applied to a synthetic APOGEE-like dataset of spectra with the objective of finding chemically identical stars without the explicit use of measured abundances.

Our work makes uses of an encoder-decoder architecture with the aim of unraveling the target parameters of the rest of the information contained in the data. Like Fader Networks, a method based on conditional autoencoders with adversarial training is proposed. This solution mixes generative network approaches that generate factors of variation through totally unsupervised methods (Chen et al. [2], Karras et al. [9]), with approaches where the aim is to obtain fully supervised representations (Hinton et al. [10]). In our case, similarly to Mathieu et al. [11], a known set of factors of variation are isolated from unknown or uncertain ones. The principal novelty of our approach is that our version of adversarial training makes use of a discriminator per factor to be unraveled. This avoids the exponential combination that occurs in the use of a single discriminator, due to the need of our domain to discretize the factor of variations, since Fader Networks are designed to deal with discrete values. Concretely, we present a method for the chemical disentanglement of the information present in stellar spectra, with application to two spectroscopic surveys, APOGEE and Gaia/RVS. The purpose is to find a projection of the stellar spectra that is encoded using latent space, and in which the contribution due to main stellar physical properties (effective temperature and surface gravity) is eliminated so as to maximize the information related to the chemical properties. Afterwards, that new representation can be used, for instance, to locate chemical peculiar objects, unconditioned by physical parameters. To test the effectiveness of our method, synthetic astronomical data from the APOGEE and Gaia/RVS surveys are applied, which present properties that favor the multipurpose testing of our method.

In addition to the original application in the field of chemical disentangling of stellar observations, the most direct outcome of our methodology is in the field of Big Data Astronomy, driven by the development of some important modern all-sky spectroscopic surveys. The proposed algorithm allows for an important data dimensionality reduction while preserving, or even highlighting, the information of interest present in them. As an added product, our method resulted in an efficient way of generating samples outside the range of the input data, which may serve as validation for the creation of new synthetic spectral grids.

GANDALF, a set of tools for training and testing our algorithm, is also provided to the community[1]. This allows for the extension of the method to different domains, as long as the input data is unidimensional. It also provides a web tool that enables us to visualize the input data, the latent space, and the reconstruction of the decoder, as a very necessary support for the follow-up and interpretation of the results. Our tool generates data that are not present in the input set, which could be used as physically consistent interpolations of the original data. Both tools are complementary, so the output of the first tool is directly usable in the visualization application.

The paper is divided into the following sections. Section 2 introduces the problem to be addressed and illustrates how the physical and chemical parameters influence the variance of a stellar spectrum, showing the case of Gaia RVS and APOGEE spectra through a grid of synthetic spectra. Section 3 describes the model or architecture for disentanglement with multi-discriminator that was developed. It also explains the approach that was followed to ensure that the disentanglement is effective, using artificial neural networks (ANN) as regressors for comparison among the different models proposed, as well as the application of the t-SNE technique (van der Maaten and Hinton [12]), a visual way to check if the disentangled parameters were indeed removed from the stellar spectra. Section 4 presents the results, displaying different tables and figures where the performance of our method is analyzed. Section 5 outlines a possible use of the proposed method for the search for peculiar stars. Finally, Section 6 summarizes the advantages that our method and our disentangling framework can offer the community within the chosen application domain, the analysis of the information contained in modern and massive multicolor sky surveys.

## 2. Data: Gaia/RVS and APOGEE synthetic spectra.

The main stellar astrophysical parameters that are used to characterize a star are its effective temperature (Teff), its gravity expressed as the logarithm of the value of the surface gravity (logg), and the atmospheric metallicity abundance [Fe/H], represented by the abundance of element Iron (Fe) in a stellar atmosphere with respect to Hydrogen. Additionally, the combination of one or several of the $\alpha$ elements (O, Ne, Mg, Si, S, Ar, Ca, and Ti) compared to iron, that is, [α/Fe], is commonly used. α elements receive their name from the fact that their most abundant isotopes are integer multiples of four – the mass of the Helium nucleus (the α particle). [Fe/H] and [α/Fe] are chemical parameters that are used in Astrophysics to assess the channels to synthesize chemical elements in stars and to study the Galactic evolution of stellar populations in the Milky Way and other galaxies [13]. Measurements of certain morphological characteristics (presence of absorption or emission lines) and their relative intensities in stellar spectra have helped astrophysicists determine astrophysical parameters and eventually establish models that represent the mass, age, and evolutionary stages of stars [14].

---

[1]https://github.com/raul-santovena/gandalf

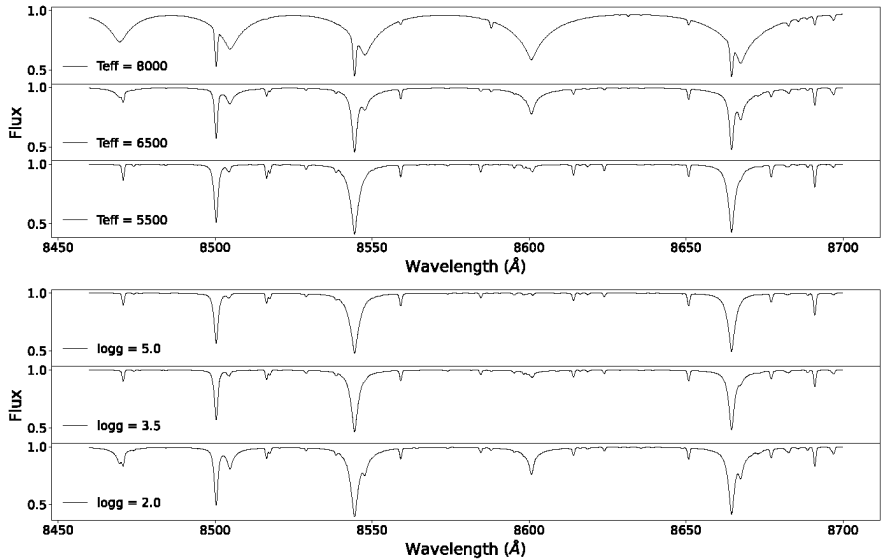

Figure 1: Example of our RVS Physical Grid: spectral flux variance for three stars as a function of [Teff] (upper panel) and logg (lower one). The values of the chemical parameters [Fe/H] and [α/Fe] are kept constant and equal to [Fe/H]=-0.75 and [α/Fe]=0.2

In this work, we shall make extensive use of the information present in two known astronomical projects. One of them, RVS, is the spectroscopic survey from Gaia mission (Gaia Collaboration et al. [15]), while the other, APOGEE (Majewski et al. [16]), is a very successful on-ground spectroscopic survey. In both cases, the use of synthetic data built from model atmospheres that mimic real observations was chosen, instead of the astronomical observations themselves.

## 2.1. Gaia mission

The European Space Agency (ESA) Gaia mission is producing a 6D, positional and kinematical map of the Milky Way, that is allowing us to go deep into our knowledge about the composition, formation, and evolution of our galaxy. Such is its importance, that it is the space mission with the greatest scientific impact in history, in terms of publications, to date. In June 2022, Data Release 3 (DR3) was published, containing about 1 million medium-resolution spectra from the RVS instrument, together with an estimation of the main stellar parameters obtained from them. Gaia satellite Radial Velocity Spectrometer (RVS) generates medium-resolution spectra (R=11500, R=$\lambda/\Delta\lambda$) in the near-infrared electromagnetic spectral region, with a range of wavelengths from 8460 to 8700 Angstrom (A), around the lines of the Calcium triplet.

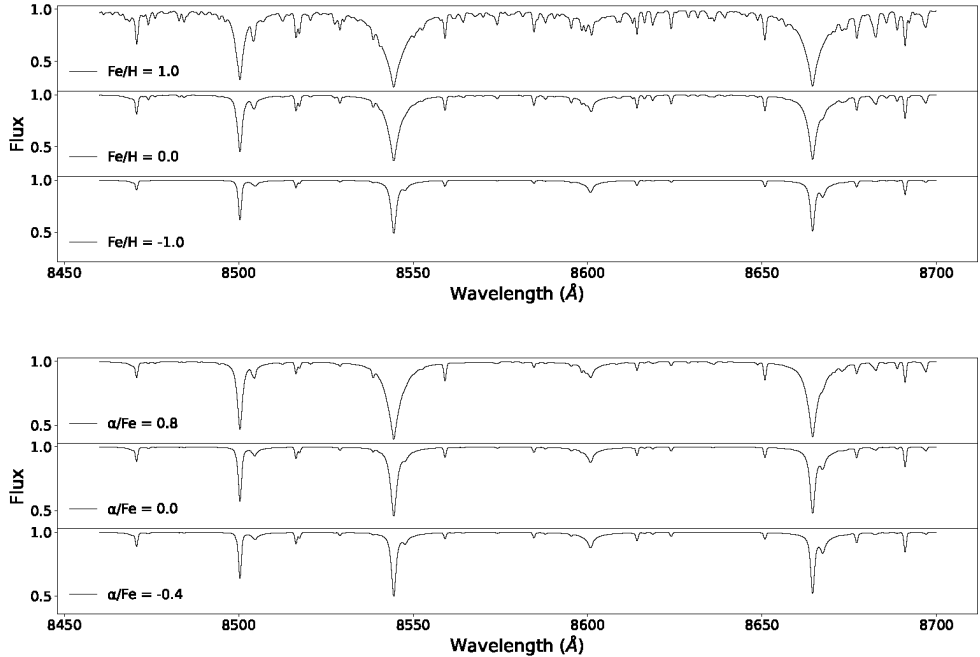

Figure 2: Example of our RVS Chemical Grid: spectral flux variance for three solar-type stars as a function of [Fe/H] (upper panel) and [$\alpha$/Fe] (lower one). The values of the physical parameters Teff and logg are kept constant and equal to Teff=6000K and logg=3.

RVS Spectra are analyzed by DPAC, the international consortium in charge of processing Gaia mission data with the objective of deriving the astrophysical parameters from the observations. The estimation of these parameters is carried out through a model-driven parametrization where the interpretation of the observed spectra is carried out through the comparison against synthetic spectra (see for instance [17] and [18]). For this purpose, DPAC computes a large grid of theoretical RVS-like spectra with different combinations of stellar astrophysical parameters, as well as chemical abundances. These spectra are calculated from MARCS atmospheric models for intermediate-mass stars using the Turbospectrum code (Plez [19]). The parameter space coverage of these grids is 2600 K to 8000 K for Teff, -0.5 to 5.5 for logg and -5.0 to 1.0 dex for the metallicity, with $\alpha$-element enrichment with respect to Iron varying between at most -0.4 dex and +0.8 dex. The computation of these grids of synthetic spectra is discussed in Recio-Blanco, A. et al. [20]. Figures 1 and 2 show several examples of the RVS grid and how the variation of effective temperature, gravity, overall metallicity, and $\alpha$-elements to Iron abundance, affects its morphology. This dataset is composed by ~38,000 spectra, with a size of 2,400 points per spectrum.

*2.2. APOGEE*

The Sloan Digital Sky Survey (SDSS) is a family of astronomical surveys, including both imaging and spectroscopy. Among its numerous projects, the Apache Point Observatory Galactic Evolution Experiment (APOGEE), is a stellar spectroscopy survey that collects spectra of Milky Way stars in the near-infrared H spectral band with a resolution of R=22,500 and a high signal-to-noise ratio ($SNR > 100$). Since SDSS IV and the arrival of APOGEE-2, the survey is covering the entire sky and the number of observations is exceeding 600,000 in its latest Data Release.

Table 1: Grid used to generate APOGEE synthetic spectra

| Parameter | Min Value | Max Value | Step |
|---|---|---|---|
| Teff | 3200 | 6000 | 200 |
| logg | 0 | 2.6 | 0.2 |

We compute a detailed grid of APOGEE-like synthetic spectra using The Payne (Ting et al. [21]). The Payne is a tool based on Kurucz stellar models and ANNs that allows accurate and precise interpolation and prediction of APOGEE spectra in high-dimensional label space, that includes both physical and chemical stellar parameters. Due to the fact that The Payne bases its approximation on data from APOGEE Data Release 14, it is a straightforward method for creating grids of synthetic data for stellar populations from the survey itself, offering flexibility in its construction. This attribute fits in a quite convenient way to the requirements of this work, helping us in the creation of a second dataset, very different from the RVS grid commented before, that serves to test our method in various scenarios.

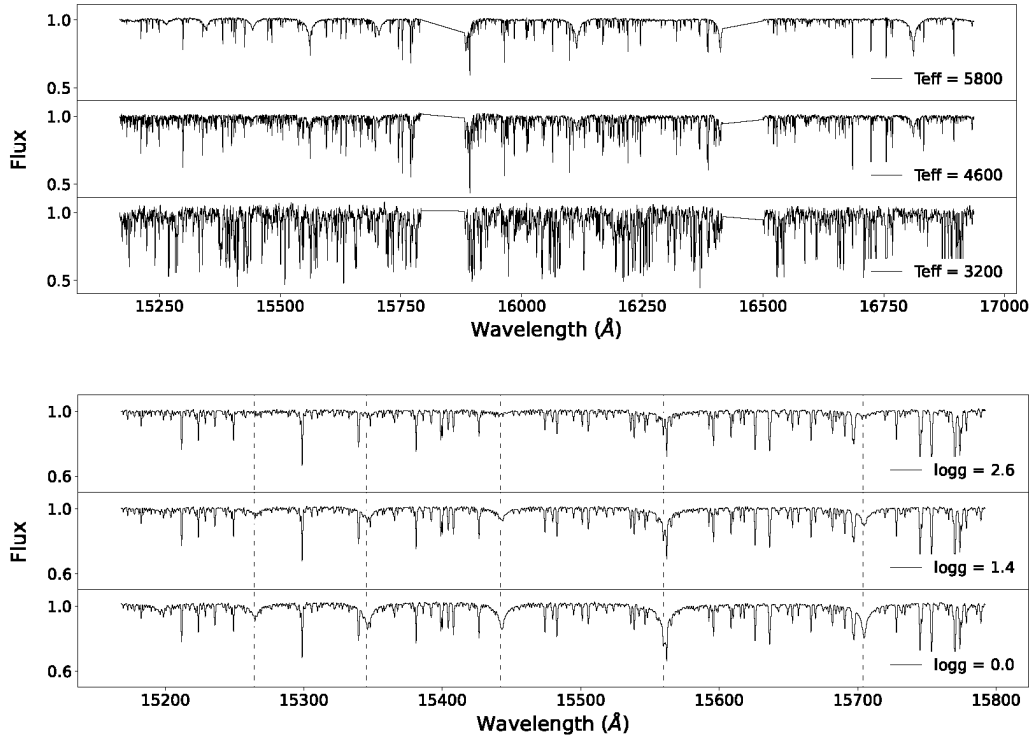

Figure 3: Examples of spectra in our APOGEE grid, showing their wavelength range as well as its richness in spectral lines. The upper figure illustrates the change in flux variance for three different effective temperatures (while keeping constant the rest of the physical and chemical parameters). The lower figure is a zoom between 15,100K and 15,800K, and it shows how the variation in logg affects the flux variance, keeping constant the rest of parameters. Some spectral lines have been pointed out with a red dashed line where greater change is perceived.

To calculate the APOGEE-like grid of synthetic spectra, the physical and chemical parameters of a random subset of 200 APOGEE real stars with good quality are selected. Because The Payne allows us to generate spectra from these chemical profiles or "fingerprints", we have generated different versions for every real star, keeping the chemistry and making variations in the temperature and gravity values. Specifically, using these 200 stars, a grid of 39,000 spectra was built, modifying their temperature and gravity according to the values shown in Table 1. In this case each spectrum consists of 7,214 points.

Figure 3 depicts some examples of APOGEE synthetic spectra as a function of effective temperature and gravity, that illustrate the morphological changes in the stellar flux when those parameters change smoothly in a star.

## 3. Method: Generative Adversarial Networks with multi-discriminators for disentangling

The objective of this work is the elimination of astrophysical parameters Teff, logg, metallicity and [$\alpha$/Fe] (hereafter conditional parameters) from the stellar spectrum. Given a spectrum $x$ and its associated conditional parameters $cp = \{cp_1, cp_2, ..., cp_n\}$, the aim is to transform the spectrum into a new representation $z$, the latent space, which does not contain information related to $cp$. In Section 4, three proposed use cases are presented in detail for comparison purposes, where different approaches that deal with all the conditional parameters, or a subset of them, are addressed.

### 3.1. Architecture

Our model is based on the use of autoencoders, an architecture formed by two artificial neural networks: the encoder $E$, which transforms the input data $x$ into a low-dimensional representation called latent space, usually written as $E(x) = z$; and the decoder $D$, which receives this abstract representation as input with the aim of reconstructing the original data ($D(z) = \hat{x}$).

In our approach, the architecture of the autoencoder is modified to force the elimination of conditional parameters (Figure 4). To achieve this, the architecture inputs these parameters twice:

- First, the encoder is modified to receive both the spectrum and the parameters as input, $E(x, cp) = z$. In this way, the necessary information associated with the spectrum is facilitated when encoding so that it can be satisfactorily eliminated (with the right training approach).

- Second, the information about the parameters is joined to the latent space as input to the decoder, $D(z, cp) = \hat{x}$. The purpose of this change is, that assuming an ideal scenario where the latent space $z$ does not contain information about the conditional parameters($cp$), it will be necessary to feed it with that information again, due to the strong influence that these parameters have on the morphology of the spectrum.

The second change consists in the use of discriminators, complementary neural networks that allow changing traditional autoencoder training for an adversarial approach, aimed at disentangling conditional parameters from latent space.

Although the model is inspired by generative models such as GANs or cGANs, it differs in some aspects such as the fact that in these methods the input of the discriminator is the output of the generator. In our case, the adversarial training is destined to the latent space, the intermediate representation of the autoencoder, and it is the input to the discriminators. The task of the generator then falls to the decoder, which uses the conditional parameters and latent space to reconstruct and generate new output data. This architecture is inspired by Fader Networks, a method proposed by Lample et al. [5] for image reconstruction from discrete attributes. We also collect some of the concepts of de Mijolla et al. [7], where Fader Networks are adapted and applied in an astronomical context, replacing images with stellar spectra, and discrete attributes by astrophysical parameters that originally present continuous values, which requires transforming them into a discretized domain. Our work provides a new approach by adding one discriminator per parameter in order to solve the combinatorial explosion that this discretization causes in the output of the discriminator when dealing with multiple conditional parameters. Although it is common and possible to address massive multi-class problems satisfactorily (see e.g. Gupta et al. [22]), this approach also has some drawbacks, such as a possible loss of precision (Moral et al. [23]) or the challenge to face problems where the number of classes compared to the number of samples is very high (Abramovich and Pensky [24]). We have therefore opted for a multi-discriminator approach, taking into account that the cost of training several discriminators represents a negligible computing time per iteration, and that the potential improvement in the performance of the discriminator can greatly help to enhance disentangling. In order to check if this improvement is real, both methods are compared in Section 4. So, to tackle the combinatorial explosion, each parameter is discretized individually in $n$ bins and a discriminator is used for each of them. The discriminators receive the latent $z$ as the input, and output a vector of size $n$ that indicates the probability of belonging to each of the bins created in the discretization process of its corresponding parameter ($p(cp^d|z)$ in Figure 4). If $n\_params$ represents the number of parameters to be disentangled, and $n\_bins$ represents the number of bins into which we divide the continuous space of each of the parameters, the output using only one discriminator will be equal to $n\_bins^{n\_params}$. Using the multi-discriminator approach where a unique discriminator is defined for each parameter, the output of each one of them will be equal to $n\_bins$, and in total there will be $n\_bins \cdot n\_discriminators$. To illustrate this, if we want to process 4 parameters using 10 bins, the original approach would have to deal with $10^4$ classes. Using multi-discriminators, there would be 10 classes for each of the 4 discriminators. Furthermore, this difference is significant considering that our datasets contain around 40,000 examples, which in the first case would leave us with a sample of 4 examples per class on average.

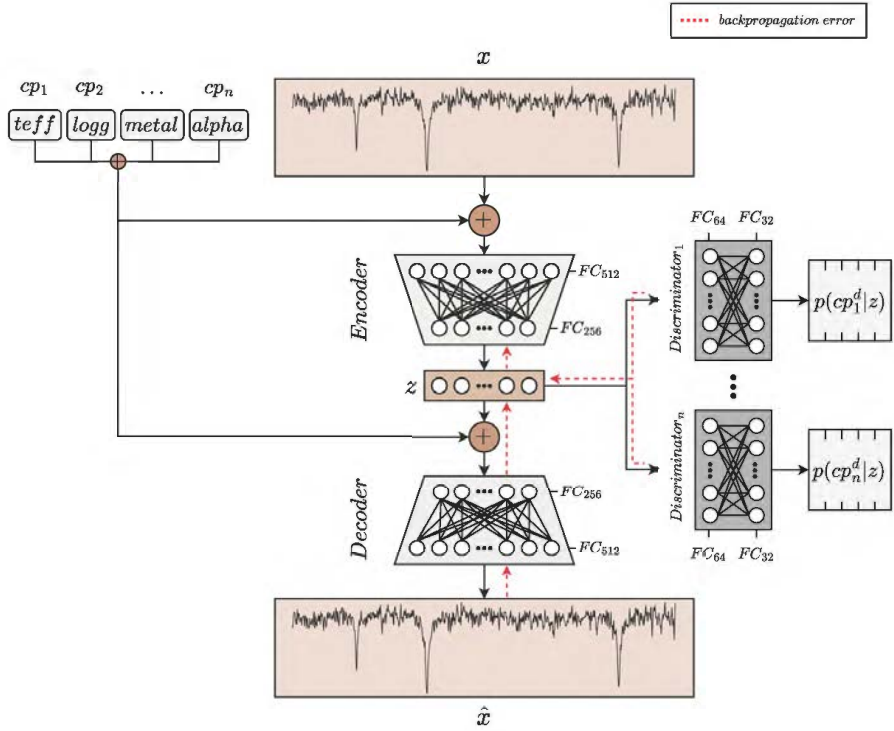

Figure 4: Disentangling architecture with multi-discriminators. Spectra $x$ and conditional parameters $cp$ are concatenate as input. The encoder maps $x \oplus cp$ to the latent space $z$, with the goal of unraveling $cp$ from $z$. The task of the decoder is to reconstruct $x$ using $z \oplus cp$ as input. Discriminators try to predict the astrophysical parameters in a discretized space $cp^d$, in such a way that their errors are used during training to teach the encoder to fool the discriminators by removing conditional parameters $cp$ from $z$. Encoder, decoder and discriminators use fully-connected neural network layers, indicated in the figure as $FC_n$, where $n$ is the number of neurons of each layer.

*3.2. Implementation*

Several tests were carried out to select the appropriate configuration of the model, using grid search techniques to find the right setups, as well as some tweaks resulting from experimentation, such as the selection of $\lambda$ values (see section 3.3 below), which respond to an analysis of the reconstruction errors versus the discriminators errors, with the aim of maintaining a balance between them. The size of the latent space was selected through a study to evaluate the relationship between its size and the information loss in the reconstruction. This has been complemented with the use of GANDALF, which allows easy and quick modification of parameters associated with both the model and the training. For the results presented in this work, the configuration is detailed below.

Based on the performed tests, the encoder, decoder, and discriminators are fully connected neural networks. The encoder is made of two hidden layers of 512 and 256 neurons respectively. Its output layer, the latent space $z$, has a size of 25 neurons. The decoder uses the inverse configuration of the encoder, with two hidden layers of 256 and 512 respectively. The input may vary because of the number of astrophysical parameters to disentangle, its size being $25 + n\_params$. All discriminators have the same configurations, with two hidden layers of 64 and 32 neurons respectively. The output size of the discriminators depends on the number of bins used in the discretization; in this case 10 bins per parameter were used.

*3.3. Training*

Following the training setup presented in Lample et al. [5], we opt for an adversarial approach, where a discriminator is pitted against the autoencoder. On the autoencoder (AE) side, it tries to reconstruct the original input through the encoding and decoding phases. For computation we use the classical mean squared error (MSE), calculated as the difference between the reconstruction of the decoder ($\tilde{x}$) and the original spectrum that serves as input to the network ($x$).

$$loss_{AE} = loss_{rec} = \frac{1}{n}\sum_{i=1}^{n}(x_i - \tilde{x}_i) \tag{1}$$

On the discriminator (DISC) side, their losses are computed using the cross-entropy or log loss, which calculate the differences between two probability distributions, the expected probability ($cp$), and the predicted probability of the discretized space ($P_{DISC}(cp^d|z)$).

$$loss_{DISC} = log\ loss = -\frac{1}{n}\sum_{i=1}^{n} y_i \cdot log P_{DISC}(cp^d|z) \tag{2}$$

Finally, the autoencoder objective is modified not only to achieve a correct reconstruction of the original data, but also to try to maximize the errors of the discriminators. In order to control the trade off between the two objectives, a factor $\lambda$ is used in the computation of the autoencoder loss. If the multi-discriminator approach is applied, a factor $\lambda$ can be defined for each of the discriminators, so that the importance given to the elimination of each parameter can be controlled.

$$loss_{AE} = loss_{rec} - \sum_{i=1}^{n_{discs}} \lambda_i \cdot loss_{DISC_i} \tag{3}$$

In retrospect, the discriminator will be trained to minimize the error in predicting the ranges of the parameters to which the latent belongs. On the other hand, the objective of the autoencoder will be to achieve a correct reconstruction of the original data, but with the new addition of trying to deceive the discriminator so that it is not capable of carrying out its objective (bigger discriminator errors will mean better disentangling). Over time, this type of training forces the autoencoder to remove from the latent space the intrinsic features of the original data that the discriminator is trying to predict, the conditional parameters.

*3.4. Assessment of disentanglement*

To test the effectiveness of our disentangling algorithm, two approximations were carried out, the first one based on the use of artificial neural networks and the second one on the t-SNE algorithm.

Our **first approach** consists of using artificial neural networks (ANNs) as regressors, which are trained to predict the astrophysical parameters. For each parameter, 3 tests are carried out by using as input data latent spaces that were generated from the following trainings:

1. No adversarial training (with $\lambda = 0$), hereafter 'noAT'.
2. Adversarial training without multi-discriminator, hereafter 'AT'.
3. Adversarial training with our multi-discriminator approach, hereafter 'mdAT'.

Those tests were compared with the results using as input the original spectra (see Section 4.1 and Table 2, for example).

The purpose of these tests is to compare the ability of a standard regressor to predict astrophysical parameters. The underlying idea is that, when the original data or a latent space generated from normal training (noAT) is used as input, the parameter predictions should be satisfactory, since in both cases there should be physical properties intrinsically encoded. On the contrary, when the input is the latent of the models that aim to eliminate the information of the astrophysical parameters (AT and mdAT), a poor prediction would be synonymous to satisfactory performance.

The $R^2$ score has been used as a metric. $R^2$ has values that vary between negatives values and 1, being the negative ones those predictions that provide arbitrary results, while 1 corresponds to perfect models that can explain the changes in the data perfectly. If the predictions are the mean of all possible values, regardless of the input, $R^2$ value will be 0. For this test, our goal for the disentangling models (AT and mdAT) will be to achieve prediction values as close to 0 as possible.

Note that several tests have been carried out with different choices of datasets for training and testing, but always maintaining the same datasets for the four cases, with the aim of making them directly comparable.

Our **second approach** is based on an unsupervised clustering algorithm, t-SNE, as a convenient way to visualize the level of disentanglement in our data. The t-distributed stochastic neighbor embedding algorithm, t-SNE, is a statistical method primarily used for visualizing high-dimensional data by giving each data point a location in a two or three-dimensional map (2 in our case). Specifically, t-SNE translates similar data to nearby areas on the 2 or 3-dimensional map. For the case at hand, the algorithm will place similar spectra near each other on a 2-dimensional map. Because astrophysical parameters themselves determine the morphology of a spectrum, if the distribution of a particular parameter in the t-SNE map is highlighted, a visual reference of the correlation between the spectrum and the parameter in question can be obtained. As it will be shown later, in this way, we take advantage of the visual power of t-SNE to provide support for our disentangling method as if it were a peer validation process.

Following the same approach as in the case of the networks, the t-SNE is applied to both the original spectra as well as to the obtained latent space. In the same way, if the proposed design method is capable of disentangling any parameter from the rest of the data contained in the spectrum, the distribution obtained after applying the t-SNE to latent spaces should not be correlated with those parameters, or at least the correlation should be less than that observed over the original dataset. This means that when we colour-code the objects according to the disentangled parameter in a t-SNE representation, we will obtain a quite random distribution of colours, indicative that such parameter has stopped conditioning the current representation of our objects. This should be consistent with the values of $R^2$, where in these cases they should be as close to zero as possible. Ideally values below 0.5 indicates a quite low correlation with the parameter, i.e. the parameter explains a small fraction of the data. Therefore, both approaches help to understand and demonstrate the degree of disentangling that were obtained.

To guarantee consistent results, the selection of t-SNE parameters has been chosen based on experimentation with various combinations. After finding a satisfactory setup, and following the author's recommendations [25], different executions have been carried out. We have used the runs with the lowest value of the objective function, in this case the KL divergence, as our final visualizations.

## 4. Results

This section explores the results achieved with the two types of tests proposed in Section 3.4. Section 4.1 studies the results obtained after applying these tests to a model trained with RVS data, and with the objective of disentangling two astrophysical parameters, Teff and logg. Section 4.2 expands the parameters to be disentangled using the RVS data again, using in this case Teff, logg, metallicity and $[\alpha/Fe]$. Section 4.3 presents the results for the APOGEE dataset using Teff and logg.

### 4.1. Results on Gaia/RVS spectra using 2 conditional parameters

Tables 2 and 3 show the results of applying the different approximations on the RVS spectra, using two conditional parameters, Teff and logg, while metallicity and $[\alpha/Fe]$ are non-conditioned. Column "X" shows the results of the prediction of the parameters from the original spectrum. The results of the column "z (noAT)" correspond to the prediction of the parameters from a latent space that is generated as a result of no adversarial training, that is, the discriminator error does not affect the training of the autoencoder ($\lambda = 0$) and therefore, does not force the elimination of the information of the conditional parameters in the latent space. This training is equivalent to working with a traditional autoencoder, but it has been done following the generator-discriminator architecture for the benefit of comparisons. Column "z (AT)" shows the results of applying the ANN regressors on the latent space of the model with adversarial training and using a single discriminator. Finally, column "z (mdAT)" corresponds to the prediction of the parameters using as input a latent space generated with an adversarial model with multi-discriminator, that is, using a discriminator for each of the parameters.

Table 2: $R^2$ results of RVS parametrization on the 2 conditioned parameters. **The lower the parametrization result, the better the disentangling.**

|  | $R^2$ score | | | |
| --- | --- | --- | --- | --- |
|  | X | z (noAT) | z (AT) | z (mdAT) |
| Teff | 0.9548 | 0.9788 | 0.8019 | 0.6240 |
| logg | 0.8773 | 0.8926 | 0.7162 | 0.2381 |

It can be seen that when the original spectra are used as input data, the prediction of both the conditioned and non-conditioned parameters is very good (first column of Tables 2 and 3, respectively), both for temperature and gravity as well as for metallicity and [α/Fe]. It is interesting to note that when the latent space is used as input, the prediction of those values, when the model is not forced to disentangle (no adversarial training, column 2 in Tables 2 and 3), is equal to, or even slightly better than that obtained from the original parameters. This means that in this case, the latent space becomes a highly compressed version of the original data, turning the originally 2400 points into only 25, that satisfactorily preserves the information. Instead, when analyzing the $R^2$ score of columns "z (AT)" and "z (mdAT)", using one or several discriminators respectively, we see how the results of the ANN worsen for the conditioned parameters, as a result of the disentangling method. Especially noteworthy is the great improvement (worsening of the predictions) that is obtained when using a discriminator per parameter, especially in the case of the logg prediction.

Table 3: $R^2$ results of RVS parametrization on the non-conditioned parameters. In this case, since these parameters are not included in the model, **high values are also expected using the mdAT model**.

|  | $R^2$ score | | | |
| --- | --- | --- | --- | --- |
|  | X | z (noAT) | z (AT) | z (mdAT) |
| [Fe/H] | 0.9511 | 0.9784 | 0.9357 | 0.9069 |
| [α/Fe] | 0.7895 | 0.8710 | 0.8494 | 0.7947 |

Table 3 allows us to verify how the discriminators affect the result of the predictions of the non-conditioned parameters, using the same dataset. These parameters are those related to the chemistry of the star, metallicity [Fe/H], and [α/Fe]. Since this data is not included in the disentangling architecture, the information on these parameters should still be contained in the data generated in the latent space.

As can be seen in Table 3, the predictions are similar in all cases, both when using the original spectra and in the different versions generated from the latent space. This indicates that although much of the information from the Teff and logg parameters is removed (Table 2), the information needed to parametrize both metallicity and α-elements overabundance is preserved.

Following the approach already described, in Fig. 5 the visualization of the dimensionality reduction performed by the t-SNE algorithm is shown. In this case, the result of applying the algorithm on the original dataset, the spectra, is compared to the one obtained when using latent space from multi-discriminator training.

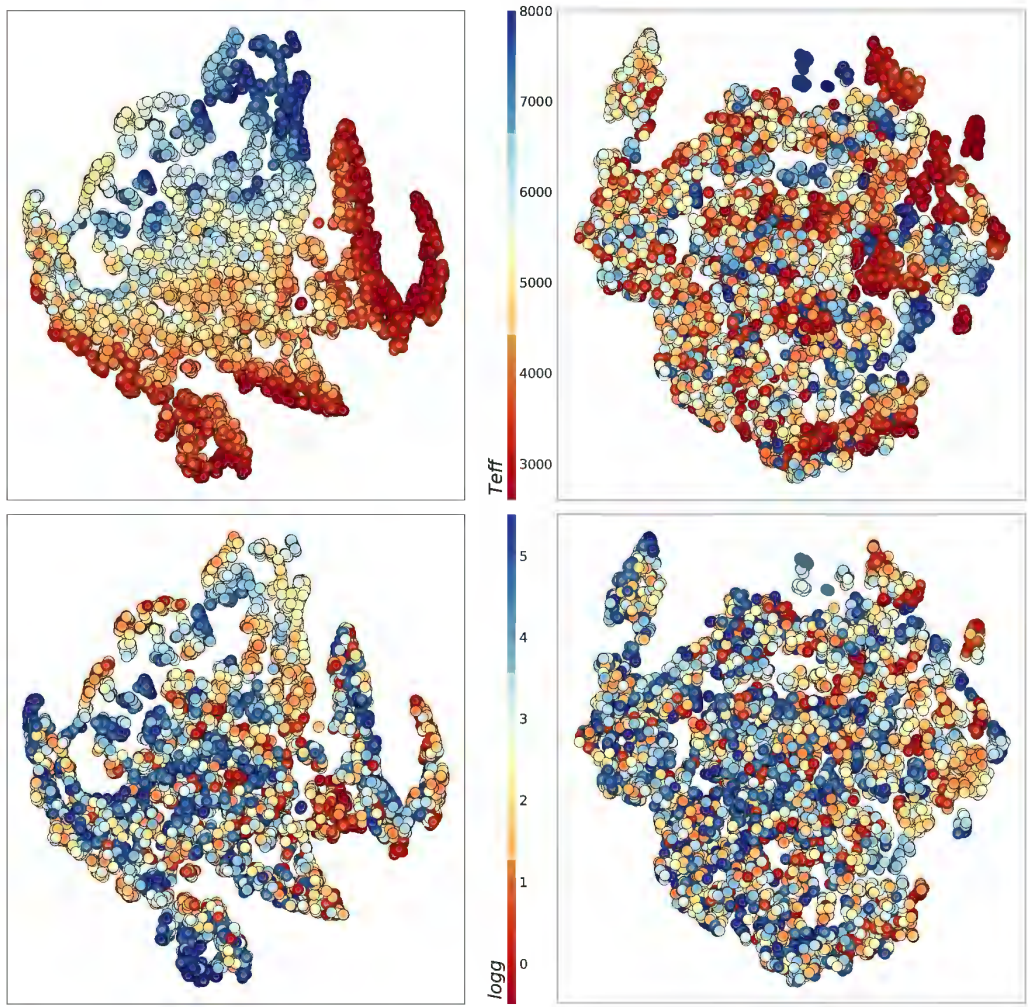

Figure 5: t-SNE clustering on the original spectra (left) and on latent space with multi-discriminator training (right). The upper row shows the data labeled according to Teff parameter, and the lower row according to logg.

The left column of the figure shows the result of applying the t-SNE to the original spectra, while the right column shows the result of applying the algorithm to the latent space. Each row refers to an astrophysical parameter (Teff in the upper row and logg in the lower one), representing each point with a color depending on the value of the parameter. For example, for the first row, the coolest stars are shown in red, while the hottest are shown in blue. As can be seen, analyzing the distribution of the data from the original spectra (left column), both rows follow a distribution with a strong dependence on the values of the parameters, more clearly appreciable in the case of temperature. Due to the fact that the t-SNE will place close-by spectra whose morphology is similar on the map, this indicates that the astrophysical parameters largely condition the morphology of the spectrum (something already known). On the other hand, from the right column (the disentangled latent space), it can be observed how this spectrum/parameter relationship is largely eliminated. These results are consistent with those obtained in the Table 2, where the prediction of temperature and gravity is almost perfect when using the original spectra, but almost random when a disentangled latent space is considered. Intuitively, the technique allows us to guide the representation obtained by the t-SNE towards the intrinsic characteristics of the spectrum of interest.

*4.2. Results on Gaia/RVS spectra using 4 conditional parameters*

Finally, a test have been made increasing the number of parameters to be disentangled up to 4 with the aim of verifying how this fact affects the results of both the architecture with one discriminator, as well as the multi-discriminator architecture. For this test, RVS data were used, applying the parameters Teff, logg, metallicity, and [$\alpha$/Fe] as conditional parameters.

Table 4 shows the obtained results. As can be seen, and applying Table 2 as a reference, the trend of the predictions is maintained when using the original spectra and the latent one. When a single discriminator training is applied, it can be seen how the predictions are still good, ergo, the disentangling model loses efficiency with 4 parameters. Also, when analyzing the results of the multi-discriminator, it can be observed how the effectiveness of disentangling is very similar to that obtained using only 2 parameters. The big improvement when unraveling 4 parameters with the multi-discriminator approach can be explained by the combinatorial explosion that occurs when working with discretized spaces and a single discriminator. In this example, by discretizing each parameter space into 10 intervals, a discriminator with 10,000 outputs is obtained. On the other hand, with the multi-discriminator, the complexity is reduced to 4 discriminators with 10 outputs each.

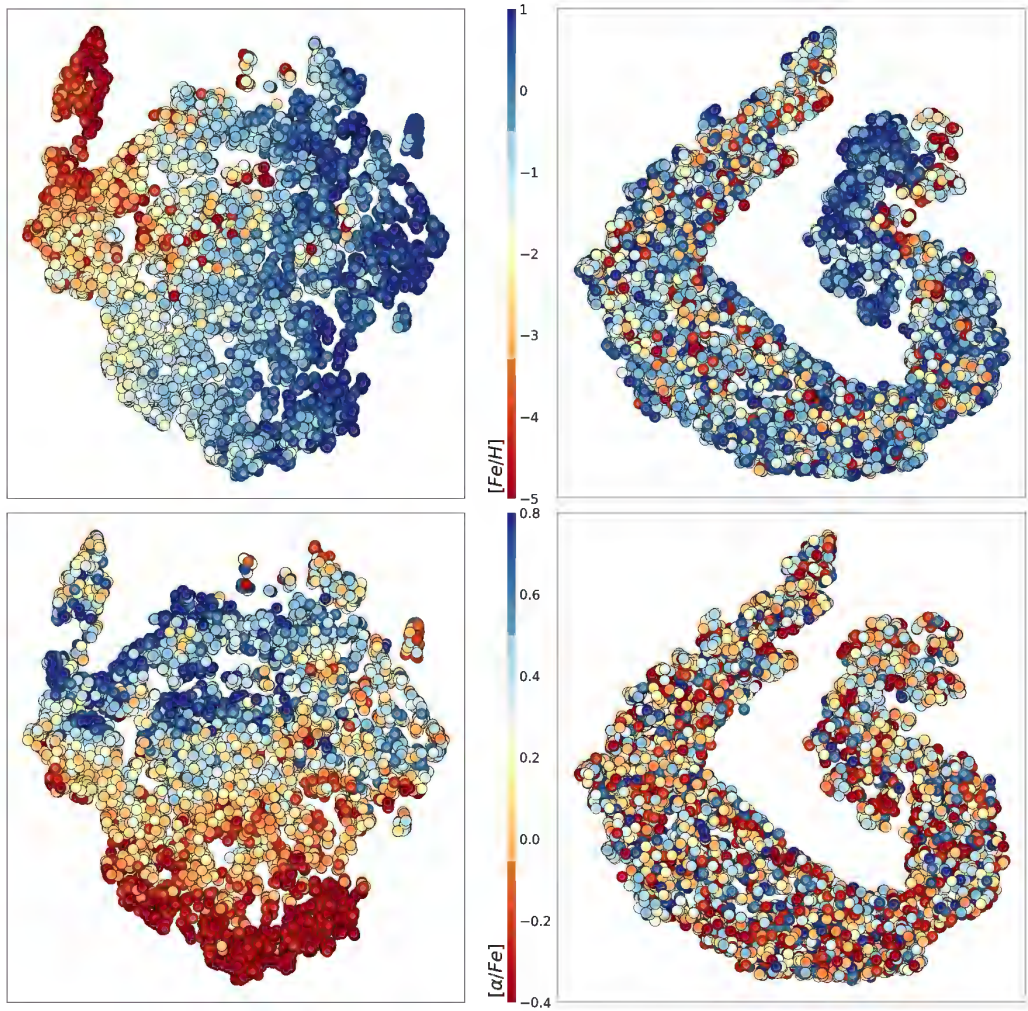

Figure 6: Comparison between metallicity and [α/Fe] when they are not disentangled as opposed to when they are. The left column represents the result of applying the t-SNE to the latent space of the model where both parameters are not, whereas the right column uses the latent space of the model where they are. Each row visualize data based on metallicity and [α/Fe] respectively.

Table 4: $R^2$ score for RVS spectra disentangling using all the parameters as conditional parameters. **The lower the parametrization result, the better the disentangling.**

|  | $R^2$ score | | | |
| --- | --- | --- | --- | --- |
|  | X | z (noAT) | z (AT) | z (mdAT) |
| Teff | 0.9400 | 0.9823 | 0.9838 | 0.5125 |
| logg | 0.8800 | 0.9255 | 0.8960 | 0.2407 |
| [Fe/H] | 0.9570 | 0.9569 | 0.9592 | 0.4854 |
| [α/Fe] | 0.7916 | 0.7326 | 0.6876 | 0.0507 |

Additionally, Figure 6 shows a comparison between the model with 2 conditional parameters in contrast to the model with 4 parameters. In this case, the figure only shows how the latent space is encoded by the t-SNE, with the intention of comparing the distributions of the metallicity and [α/Fe], which were not disentangled in the first model. Despite the fact that the result of applying the t-SNE to the latent space when performing a 4-parameter disentangling shows a quite singular distribution, it can be clearly seen how the distribution of parameters in the sample changes completely, disordering the objects and preventing the result from being conditioned by metallicity and [α/Fe].

Finally, using the model with 4 parameters and [α/Fe] as a reference, Figure 7 shows the different distributions after applying the t-SNE to the 4 possible scenarios: "X", "z (noAT)", "z (AT)" and "z (mdAT)". The obtained results are in line with what is shown in Table 4 (last row), and serve as a demonstration of the great improvement obtained in the multi-discriminator approach when the number of conditional parameters is increased.

*4.3. Results on APOGEE spectra*

The results of the previous disentanglement procedure applied to APOGEE spectra are discussed next. Table 5 shows the results of the predictions of the regressive ANNs from the original spectra, and from the latent spaces obtained from the 3 types of training mentioned in section 3.4.

As can be seen, the results follow a similar trend to those analyzed with the RVS dataset. While very good temperature and gravity predictions are obtained from the original spectra as well as with the latent space without adversarial training, when disentanglement is applied the prediction drops drastically, especially with multi-discriminator training. In this case, the elimination of the spectrum/parameter dependency is even greater than in the case of RVS spectra, a fact supported by the results of the t-SNE algorithm.

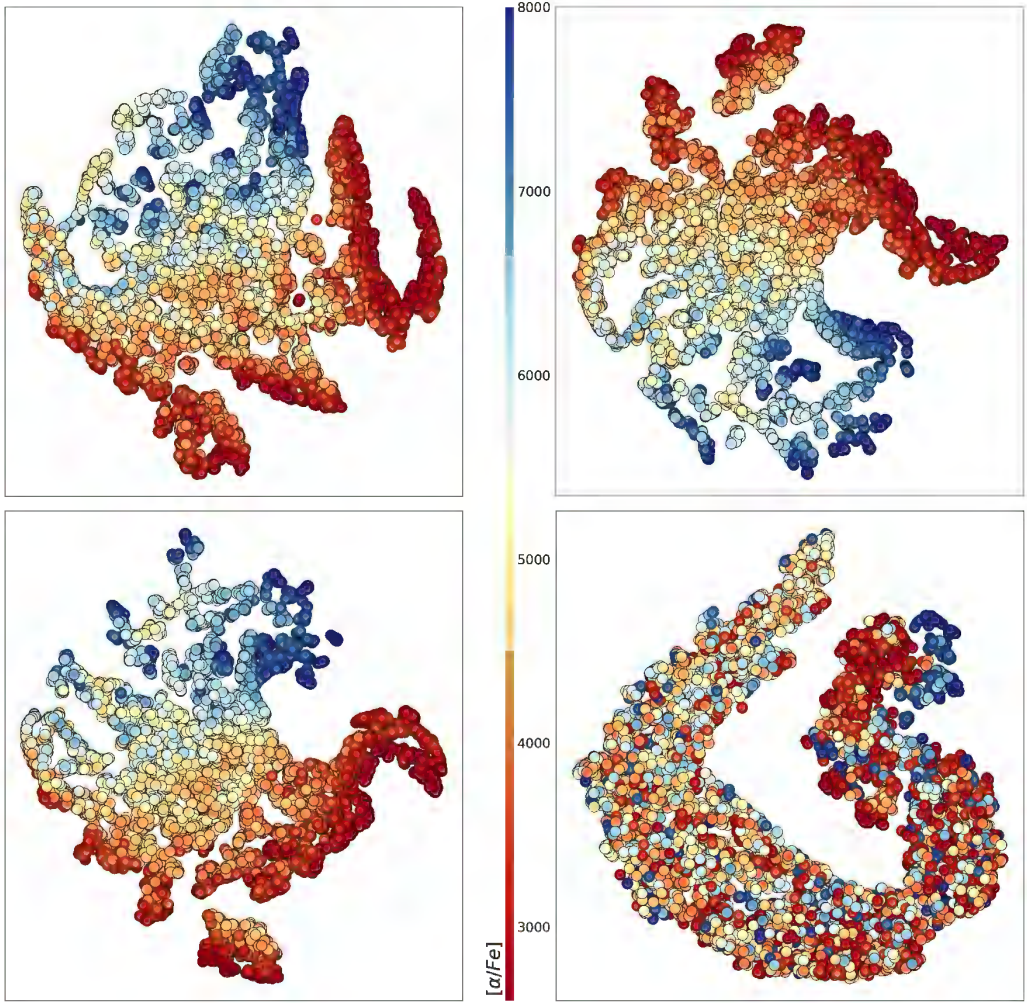

Figure 7: t-SNE distribution depending on [α/Fe] values and using as input (from left to right and top to bottom): "X", "z (noAT)", "z (AT)" and "z (mdAT)".

Table 5: $R^2$ results of APOGEE spectra parametrization on the conditioned physical parameters. **The lower the parametrization result, the better the disentangling**.

|      | $R^2$ score | | | |
|------|--------|-----------|--------|----------|
|      | X      | z (noAT)  | z (AT) | z (mdAT) |
| Teff | 0.9968 | 0.9899    | 0.8910 | 0.2244   |
| logg | 0.9851 | 0.8192    | 0.5897 | 0.1436   |

Figure 8 shows a comparison of the t-SNE application to the original spectra and the latent space generated after training with a multi-discriminator. As mentioned, it can be clearly seen how the astrophysical parameters totally condition the distribution obtained from the original data (graphs in the left column). This effect is even more evident in the case of APOGEE than in that of RVS spectra, as can be seen in the upper left graph where the temperature values are represented. On the other hand, analyzing the result of the latent space, the clustering obtained by applying the t-SNE shows almost a total independence of both parameters, Teff and logg. The significant differences between the results obtained with the RVS and APOGEE spectra are due to the fact that the APOGEE spectra cover a broader spectral range and have better wavelength resolution. In consequence they contain more information related to the astrophysical parameters, allowing the model to better understand the dependence between the conditional parameters and the spectrum.

## 5. Discussion

This section analyze the importance and meaning of the results obtained beyond the metrics presented. For this, the APOGEE results have been chosen. On the one hand, they contain greater detail in the spectra and present better results. On the other hand, the initial motivation to implement a disentangling method arose from the idea of utilizing this technique to seek out stars with similar characteristics in large astronomical databases, and APOGEE has been widely used in that regard.

In a previous work focused on the discovery of Phosphorus-rich stars (Masseron et al. [26]), we show that fast methods based on signal processing techniques can be very useful for the search for peculiar stars, as they can be adapted to the particularities of any spectral database. These methods explore the spectrum in search of spectral lines, which arise as a result of radiation transfer in the stellar photosphere, where different atoms and molecules exist and give rise to spectral lines at specific wavelength intervals.

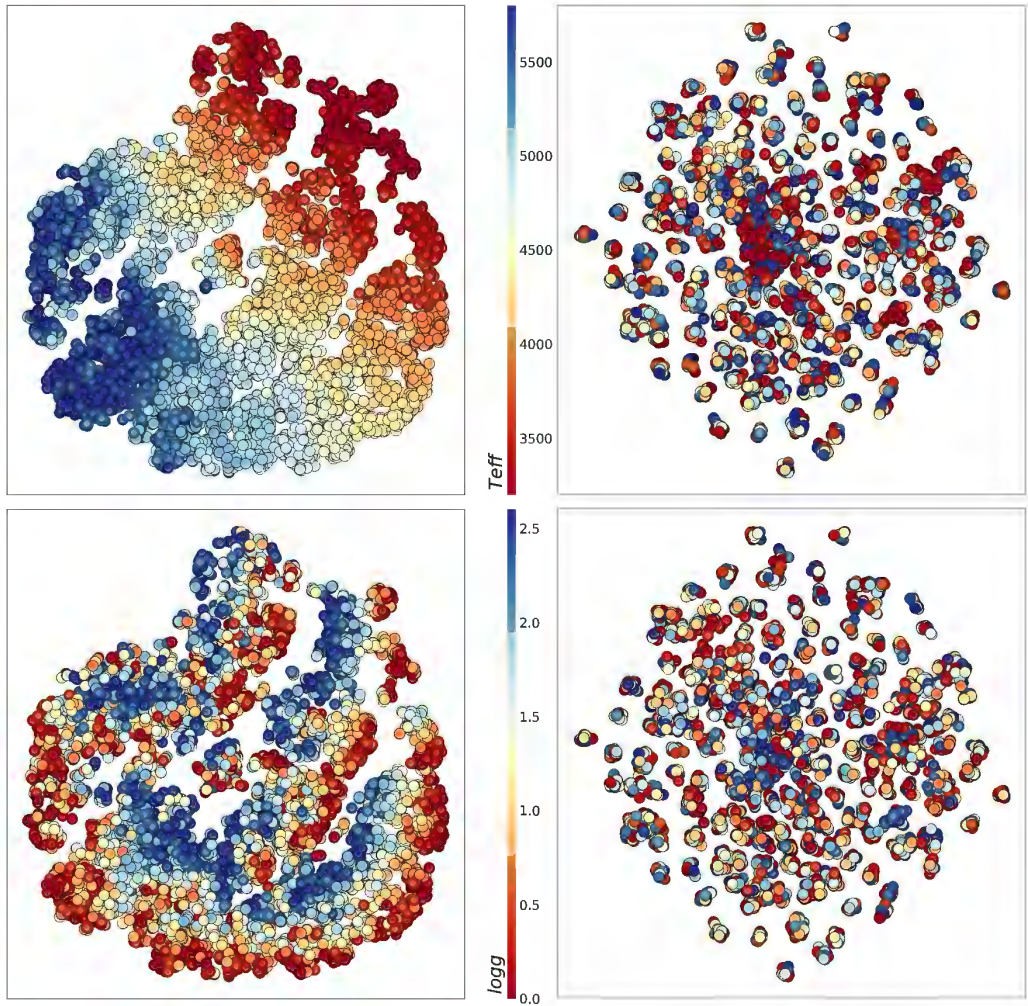

Figure 8: t-SNE application to APOGEE using the original spectra as input (left column), and using the latent space obtained from a multi-discriminator model (right column). The top row shows data colored based on temperature, while the bottom column shows data based on gravity.

However, these approaches are strongly affected by the physical conditions in the stellar environment, which greatly affect the presence or absence of specific lines as well as their morphology and intensity as illustrated in Figure 3. This interdependence of physical and chemical factors makes it difficult to detect the lines, measure their intensities, and determine the existence of any peculiarity such as an overabundance of a particular chemical element. Another condition is the fact that in many cases, there is no prior sample of what you want to search for, or it is very small, precisely because of its rarity. In this case, any supervised method is also useless. If an unsupervised method is chosen, it is again necessary to deal with the fact that any method that takes the original spectra as input would not be free from the influence of physical conditions on the spectral morphology.

We believe that disentangling is an alternative that can deal with the problems described above, since it can be used to eliminate information that may affect the search for these stars. However, it is important to highlight that the result of disentangling is an abstract representation of the spectrum itself, the latent space, that cannot be used to search for specific spectral lines related to the presence of a particular chemical element. Due to this, other alternatives must be explored to be able to work with this new representation. The use of clustering methods such as the classical k-means, or dimensionality reduction algorithms such as t-SNE or PCA allows an approach focused on the search for similarities in the latent space.

Therefore, we delve into the significance of our results and their potential application in the search for peculiar stars, exemplifying that the generation of a latent space partially free of astrophysical parameters can help their detection. To do this, Figure 9 again illustrates the comparison of the results obtained in section 4.3 after the application of t-SNE in APOGEE, but this time showing the distribution of abundance of two chemical elements, Manganese and Vanadium, before and after disentangling. In the two cases presented, it can be seen that after disentangling it is possible to group stars that share chemical features using their latent, which establishes a first step for their detection through disentangling. It also should be noted, as seen in the figure, that the distribution of the latent based on chemistry continues to present intermixed areas. One of the reasons, as reflected in the results tables of section 4, is that disentangling is not perfect, therefore there is residual temperature and gravity information that continues to condition the distribution; and second, that the latent is also conditioned by the abundances of all the elements that make up the chemical profile of the star, which makes it very difficult to separate all those elements in a low-dimensional space. For this reason, it is interesting to also analyze the distribution of latent space not only for individual abundances, but also as a whole. The APOGEE dataset was created precisely by randomly selecting 200 stars from the database, each of them with its own chemical profile or "fingerprint", that is, with specific values of elementary chemical abundances. By varying the astrophysical parameters Teff and logg of each one, we then generated different versions of synthetic spectra that preserve the original chemical profile. Based on this, Figure 10 shows the distribution of those 200 chemical profiles originally used, although for the sake of clarity, only 20 are colored. Analyzing the figures together, two immediate conclusions can be drawn:

i) First, the method is capable of generating similar latents for stars with identical chemical profiles, despite the fact that they present large variations in their physical parameters that greatly alter their morphology (as observed in Figure 3).
ii) Second, it is also capable of constructing similar latents for different stars but that share overabundances in a specific element. Based on the union of both characteristics, we believe that the combination of disentangling together with unsupervised methods with similarity metrics can be a real application in star detection.

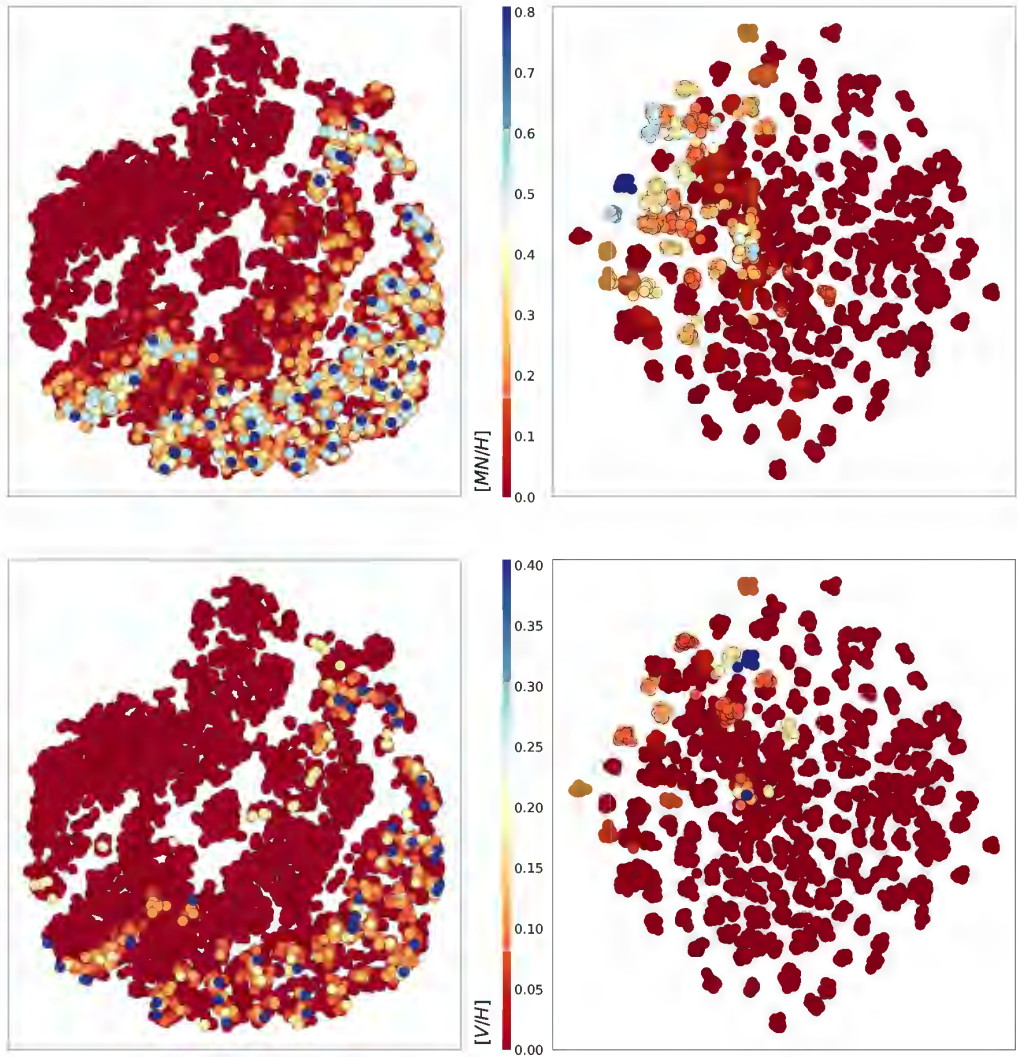

Figure 9: Cluster selection of t-SNE visualization in APOGEE, left panel for the original data and right panel for the latent space, where Teff and logg are disentangled. The sidebar reflects the values of the elementary abundances, Manganese with respect to Hydrogen abundance ([Mn/H])) in the upper panels and Vanadium with respect to Hydrogen ([V/H]) in the lower ones. A cutoff for values below 0 (less than the sun) has been applied to highlight overabundances (colored in dark blue).

Nonetheless, it is important to highlight that this methodological approach, which seems very promising to automate the search for chemically peculiar objects, must still be developed and tested in detail for specific cases of astrophysical interest. This matter lies beyond the scope of the present paper.

Finally, despite its potential uses in this field, there are also drawbacks that must be discussed. The latent space is nothing more than a compressed representation of the astronomical spectrum whose relationship with the parameters to be measured cannot be directly interpreted. Fully supervised approaches can be considered with the idea of fully encoding the factors of variation, but in this case they would require precise data on the chemical parameters of the stars. In astronomical sky surveys, the quality of physical and chemical information is subject to multiple factors which can greatly complicate their interpretation. On the other hand, the difficulty of separating the factors of variation increases with observational datasets, motivated again by elements such as noise from the environment or instrumentation. As previously mentioned, other methods should also be considered for searching for similar data based on the latent beyond the t-SNE, which at this

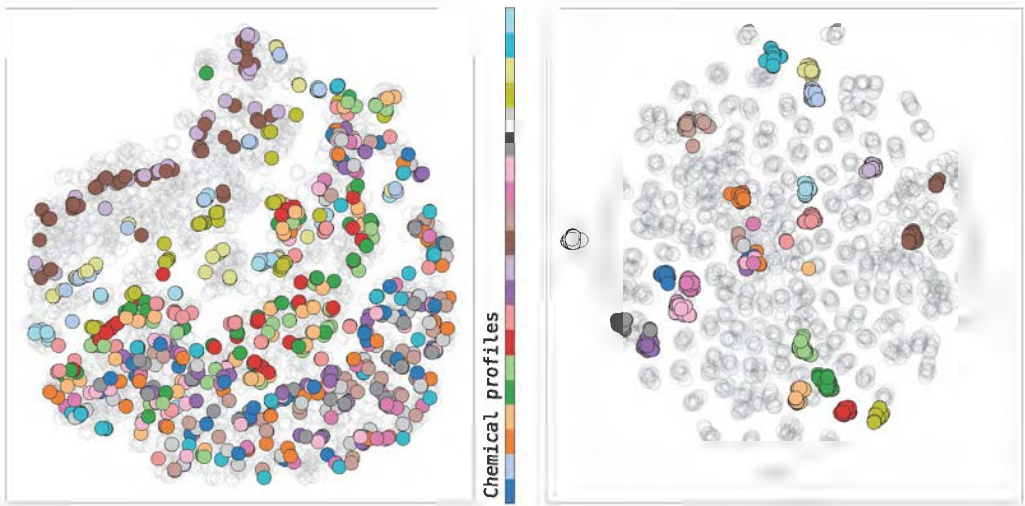

Figure 10: Cluster selection of t-SNE visualization in APOGEE, left panel for the original data and right panel for the latent space. We illustrate the distribution based on their chemical profiles. Each color represents the unique set of individual abundance values. The different points for each color represent the several variations made by modifying the astrophysical parameters.

To enhance visualization, a random subset of 20 profiles has been selected; the remaining profiles are shown in gray.

point is merely illustrative. From the perspective of disentangling, taking into account alternatives that establish restrictions in the construction of the representation can also be very useful. For example, the use of Variational Autoencoders (Kingma and Welling [27]) that better exploit the parameter space of the latent space can indirectly favor the subsequent analysis of this data, also considering that Variation models in combination with generative networks can help improve the generative part of the process (e.g. Razghandi et al. [28].).

There are many paths to explore before setting a well-founded framework for its direct application, but based on the results obtained, we believe that this methodological approach should be considered and studied in more detail in future works.

## 6. Conclusions and future work

For the first time, our version of adversarial training makes use of a discriminator per parameter to be unraveled, thus avoiding the exponential combination that occurs in the use of a single discriminator due to the discretization of the values to be disentangled. While the proposed technique can be applied in various fields, this article specifically demonstrates its applicability in astronomy for the disentanglement of information in two spectroscopic surveys: APOGEE and Gaia/RVS.

In addition to the original application, the most direct outcome of our methodology is in the field of Big Data Astronomy, driven by the development of a number of important modern all-sky spectroscopic surveys. Our algorithm provides an important data dimensionality reduction while preserving, and even highlighting, the information of interest present in them. As an added product, our method resulted in an efficient way of interpolating and extrapolating samples outside the range of the input data, which may serve either as validation or for the creation of new synthetic spectral grids.

To carry out this work, an ad-hoc framework called GANDALF was built. It is composed of several Python classes to generate data, command line tools to define, train, and test the models; and a web application to show how the algorithm works in a visual way. GANDALF is available for the community as an open source tool hosted in Github. This allows the community to replicate the presented results and use and adapt disentangling to their field of work. It also takes advantage of the generative models by offering the option of creating new spectra with astrophysical parameters not included in the definition of the original synthetic grids. In short, this work presents some substantial improvements to representation learning techniques presently available in the literature, and explores the use of t-SNE and ANN as tools to asses the quality of the disentanglement that has been obtained. Our approach is supported by the use of synthetic stellar spectra datasets. This favors a study focused on models, avoiding the influence of external factors that could alter the morphology of the spectra, such as interstellar dust, nearby objects, or calibration errors in the measuring devices. Finally, this work also provides support for the community, offering a demonstrative framework to apply these techniques in this and other fields such as signal processing, where these techniques could be applicable.

This work leaves several lines open for the future. Although the use of synthetic datasets is the best choice to perform qualitative analyzes of our approach, it is necessary to consider the use of real data. In this context, experimental tests are currently being carried out with the GANDALF framework to improve the parametrization of stellar parameters that DPAC performs with RVS spectra on the Gaia mission. As discussed above, another future line lies in the exploitation of latent space for the detection of peculiar stars. In view of the results presented in Section 5 using the t-SNE algorithm, the use of unsupervised approaches that explore latent similarities represents a promising field of work. In this case, to avoid dealing with the aforementioned drawbacks of working with real spectra, it is believed that a mixed approach that uses real information to generate synthetic spectra can be beneficial. Lastly, although GANDALF is intended as a multi-purpose framework, it is currently only compatible with one-dimensional data. Our team is working on generalizing it so as to work with multi-dimensional data.

## 7. Acknowledgments

Horizon Europe funded this research [HORIZON-CL4-2023-SPACE-01-71] SPACIOUS project, Grant Agreement no. 101135205, the Spanish Ministry of Science MCIN / AEI / 10.13039 / 501100011033, and the European Union FEDER through the coordinated grant PID2021-122842OB-C22. We also acknowledge support from the Xunta de Galicia and the European Union (FEDER Galicia 2021-2027 Program) Ref. ED431B 2024/21, CITIC ED431G 2023/01, and the European Social Fund - ESF scholarship ED481A2019/155. M.M. acknowledges support from the cooperation agreement between the IAC and the Fundación Jesús Serra for visiting grants.